\pdfoutput=1
\documentclass{article}
\usepackage{spconf,amsmath,graphicx}
\usepackage{threeparttable,booktabs,multirow, tabularx}
\usepackage[caption=false]{subfig}

\begin{document}
%

\title{Unsupervised Anomaly Detection and Localization of Machine Audio:\\A GAN-based Approach}

\name{Anbai Jiang$^{1}$, Wei-Qiang Zhang$^{1}$, Yufeng Deng$^{1}$, Pingyi Fan$^{1*}$, Jia Liu$^{1,2}$
\thanks{*Corresponding Author\newline \indent \indent This work was supported by the National Key Research and Development Program of China (Grant NO.2021YFA1000504) and the National Natural Science Foundation of China under Grant No. 62276153.}}
\address{$^{1}$BNRist, Department of Electronic Engineering, Tsinghua University, Beijing, China\\
$^{2}$Tsinghua AI Plus, Beijing, China}

\maketitle
\begin{abstract}
Automatic detection of machine anomaly remains challenging for machine learning. We believe the capability of generative adversarial network (GAN) suits the need of machine audio anomaly detection, yet rarely has this been investigated by previous work. In this paper, we propose AEGAN-AD, a totally unsupervised approach in which the generator (also an autoencoder) is trained to reconstruct input spectrograms. It is pointed out that the denoising nature of reconstruction deprecates its capacity. Thus, the discriminator is redesigned to aid the generator during both training stage and detection stage. The performance of AEGAN-AD on the dataset of DCASE 2022 Challenge TASK 2 demonstrates the state-of-the-art result on five machine types. A novel anomaly localization method is also investigated. Source code available at: \underline{www.github.com/jianganbai/AEGAN-AD}
\end{abstract}
\begin{keywords}
Anomaly Detection, Machine Audio, Autoencoder, Generative Adversarial Network
\end{keywords}
\section{Introduction}
\label{sec:intro}

Anomaly detection aims to identify anomalous samples from normal samples when only normal samples are provided. With the development of modern manufacturing, there has been an increasing need for machine anomaly detection, especially via audio. Recent years saw more and more effort devoted to anomaly detection via image or sensor signal. However, audio-based anomaly detection remains challenging for both conventional models and deep learning based models. There are the following difficulties to be overcome:

\begin{enumerate}
\item[(1)]\textbf{Paradox in feature representation: non-intuitive or high dimensional.} It is well-known that audio carries message in a time-frequency coupling form, whose raw waveform does even make sense for humans. To present it in an intuitive form, spectrogram is calculated on the time series, yielding a high dimensional representation which exceeds the detection capacity of conventional algorithms, including k-nearest neighbor (KNN) \cite{ramaswamy2000efficient} and local outlier factor (LOF) \cite{breunig2000lof}. 
\item[(2)]\textbf{Indirect and Unreliable Supervision.} Deep learning has demonstrated its capability of handling high dimensional input in the past few years. However, concerning anomaly detection, challenges remain since deep learning based models must learn to dig out hidden patterns through the enormous input without direct supervision. Classification model \cite{LiuCQUPT2022}, averts this restriction by training on auxiliary labels, yet there is no guarantee that this yields a good feature extractor.
\item[(3)]\textbf{Variational Working Conditions.} The variation of working conditions leads to distinct feature representation, making it harder for models to identify between anomalous clips and normal clips from rare working conditions, especially when similar patches can be seen in both normal and anomalous distributions.
\end{enumerate}

Considering these problems, GAN \cite{goodfellow2020generative} may be a potential breakthrough for audio anomaly detection, not only because it can be trained without utilizing auxiliary labels, but also because of its great potential of learning huge manifold and generalizing on multiple domains. Recent advances has manifested GAN's effectiveness on image anomaly detection, including AnoGAN \cite{schlegl2017unsupervised} which searches the optimal latent variable and GANomaly \cite{akcay2018ganomaly} where the generator is modified as an autoencoder. However, it is believed that AnoGAN is not suitable for audio anomaly detection, since the normal and anomalous distributions of spectrogram have minor differences, which contradicts its basic assumption. Reconstruction seems to be a more feasible approach, as reconstruction models have already been proposed in DCASE Challenge. Typical reconstruction models \cite{DuNERCSLIP2022, YamashitaGU2022} are trained to reconstruct spectrogram and the reconstruction error is utilized as an indicator of anomaly. Ordinary explanation is that the reconstruction error is expected to be bigger for anomaly since the model is only trained on normal samples. 

\begin{figure}[ht]
    \subfloat[]{
        \includegraphics[width=0.45\linewidth]{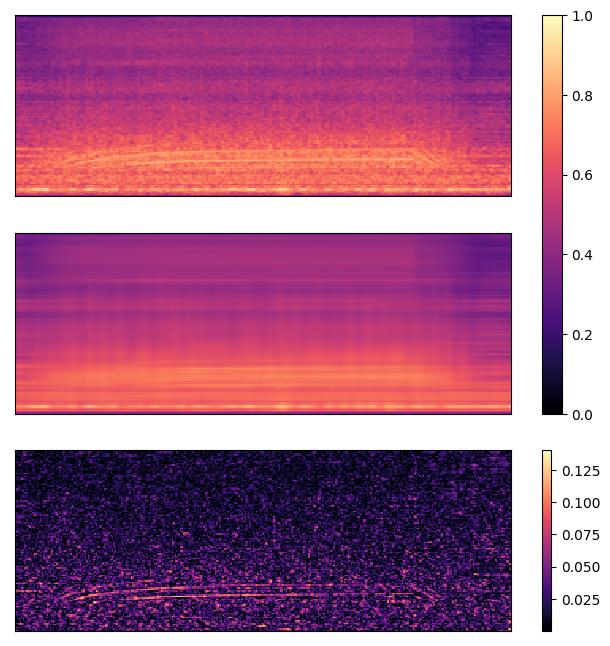}
        \label{fig:aerethink_expect}}
    \hfill
    \subfloat[]{
        \includegraphics[width=0.45\linewidth]{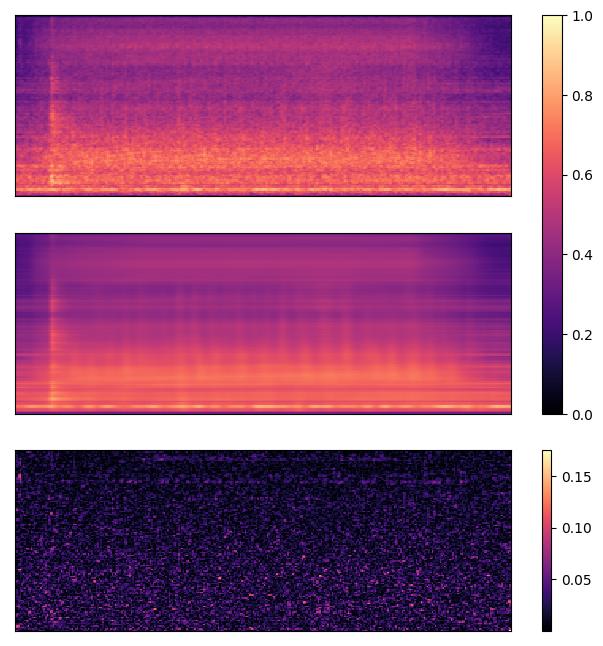}
        \label{fig:aerethink_most}}
    \caption{Reconstruction result of two anomalous ToyCar spectrogram by an autoencoder. First row: Original. Second row: Reconstructed. Third row: Absolute error. (a) A typical demonstration of the ordinary explanation with distinct residual, yet few clips belong to this pattern. (b) The reconstruction error of most clips is plain noise. For this clip, the abrupt impulse is still well-reconstructed.}
    \label{fig:aerethink}
\end{figure}

However, we argue that reconstruction model only learns \textbf{how to denoise properly} when the normal and anomalous distributions are mostly overlapped, rather than learning the whole distribution (i.e. what should appear and what shouldn't). Trained by mean square error (MSE), reconstruction model more or less resembles principal component analysis (PCA). In that way, the model only learns how to preserve main components and how to discard random noise. It is observed in Fig \ref{fig:aerethink} that the anomalous components of most anomalous spectrograms are still well-reconstructed by an autoencoder, which is contrary to the ordinary explanation. Reconstruction error is mostly contributed by random noise, making it an invalid indicator of anomaly. Therefore, a discriminator is introduced to provide a feature-level guidance  for the autoencoder (also the generator), aiming to enforce the autoencoder to dive into the representation rather than focusing on the superficial noise. The discriminator and the generator are adversarially trained by WGAN-GP \cite{gulrajani2017improved} and feature matching loss \cite{salimans2016improved} respectively. AEGAN-AD detects anomaly from two complementary perspectives: reconstruction error based on the generator and embedding extracted from the discriminator which is then processed by conventional detection algorithms. We also explore the potential of localizing anomalies by comparing the reconstructed (denoised) spectrogram and the mean representation, which can be of great help for smart diagnosis.

\begin{figure}[ht]
    \centering
    \includegraphics[width=0.95\linewidth]{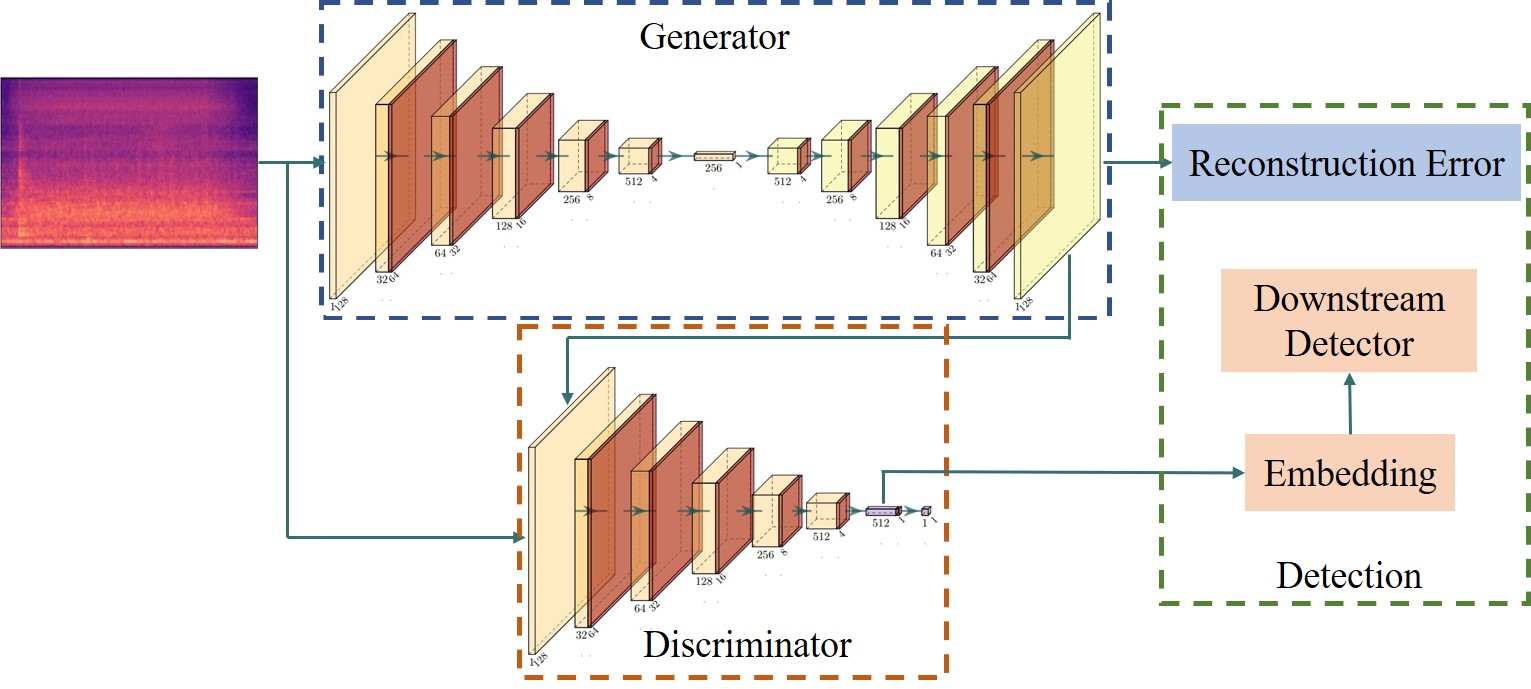}
    \caption{The training and detection procedure of AEGAN-AD.}
    \label{fig:framework}
\end{figure}

The experiment is conducted on the dataset of DCASE 2022 Challenge TASK 2 \cite{MIMII, ToyADMOS}, compared with two official baselines and five generative models, over five machine types (bearing, fan, gearbox, slider, ToyCar). Our AEGAN-AD model presents the state-of-the-art performance among unsupervised models, surpassing seven baselines with a general improvement of 3.84\% in terms of the area under the receiver operating characteristic (ROC) curve (AUC) and partial-AUC (pAUC). The ensemble model \cite{deng2022ens}, which comprises AEGAN-AD and other models, ranked No.1 on development set and No.4 on evaluation set in DCASE 2022 Challenge.

\section{Proposed Method}
\label{sec:method}
A general framework of AEGAN-AD is illustrated in Fig \ref{fig:framework}.
\subsection{Model Setup}
In the proposed AEGAN-AD model, the generator \textit{G} is modified as an autoencoder which reconstructs mel-sepctrogram, while the discriminator \textit{D} is trained to discriminate between real spectrograms and reconstructed spectrograms. The structure of \textit{G} can be viewed as an inverse cascade of deep convolutional GAN (DCGAN) \cite{radford2015unsupervised}, with the encoder of \textit{G} being the discriminator of DCGAN and the decoder of \textit{G} being the generator of DCGAN. The structure of \textit{D} mainly resembles the encoder of \textit{G}, except that the last layer is substituted by a depth-wise convolution to promote the extraction of semantic embedding. Layer normalization (LN) \cite{ba2016layer} is mainly adopted so that statistics of target domain will not be covered up by that of source domain, which preserves distinguishable features and promotes the scalablity on both domains.

\begin{table*}[ht]
    \centering
    \caption{Performance Comparison with Baseline Systems}
    \label{tab:comp_related}
    \renewcommand{\arraystretch}{1.1}
    \begin{tabularx}{\linewidth}{ccccccccc}
        \toprule
        \makebox[0.1\linewidth][c]{Machine Type} & \makebox[0.08\linewidth][c]{Official-1} & \makebox[0.08\linewidth][c]{Official-2} & \makebox[0.08\linewidth][c]{AnoGAN} & \makebox[0.08\linewidth][c]{GANomaly} & \makebox[0.08\linewidth][c]{Du} & \makebox[0.08\linewidth][c]{To\v{z}i\v{c}ka} & 
        \makebox[0.08\linewidth][c]{Yamashita} & \makebox[0.08\linewidth][c]{Ours} \\
        \midrule
        bearing & 54.80 & 59.18 & 52.70 & 55.40 & 59.80 & 65.12 & 60.24 & \textbf{76.03} \\
        fan & 58.47 & 57.20 & 41.89 & 53.60 & 54.16 & 61.46 & 62.37 & \textbf{65.83} \\
        gearbox & 63.07 & 59.89 & 49.45 & 53.44 & 68.40 & 74.08 & 70.62 & \textbf{75.27} \\
        slider & 57.99 & 50.17 & 48.77 & 63.10 & 69.90 & 71.52 & 72.15 & \textbf{74.06} \\
        ToyCar & 51.06 & 54.26 & 41.41 & 52.48 & 64.93 & \textbf{79.94} & 69.25 & 78.46 \\
        hmean & 56.80 & 55.90 & 46.42 & 55.36 & 62.88 & 69.82 & 66.58 & \textbf{73.66} \\
        \bottomrule
        \multicolumn{9}{p{\textwidth}}{Official-1 is the dense autoencoder baseline provided by DCASE organizers which reconstructs spectrogram segment. {\par}Official-2 is the classification baseline provided by DCASE organizers. hmean is the harmonic mean over all machine types.}
    \end{tabularx}
\end{table*}

The discriminator of AEGAN-AD is trained by WGAN-GP, while the loss function of the generator is modified as feature matching loss, a constraint on sample level. This is because supervision on distribution level brings ambiguity for anomaly detection. In that way, samples will be processed to resemble the whole distribution and reconstruction error will be equally big for both normal and anomalous samples. Based on this thinking, we extract embedding from the discriminator and compare the mean and the standard deviation of the embedding from the real distribution with that of the reconstructed distribution. The overall loss function of the generator combines the difference in statistics with MSE. The first term is expected to reduce the tendency of denoising and promote deep understanding of input spectrogram, while MSE is still reserved to improve the robustness. 

\subsection{Detection Procedure}
Anomalies are detected from two complementary perspectives: reconstruction error measured by the generator and embedding extracted from the discriminator. As for the generator, reconstruction error is measured in both sample space and latent space. The first part compares the real spectrogram with the reconstructed spectrogram. The second part compares the latent representation of the real sample with the latent representation of the reconstructed sample, since the latent space highlights key information of each sample. L2 norm, L1 norm and cosine distance are adopted to measure the difference and each measurement corresponds to an anomaly score.

On the other hand, embedding extracted from the depth-wise convolution layer of the discriminator is expected to possess rich semantic features. Therefore, the embedding is processed by conventional anomaly detection algorithms including KNN, LOF and distance measurements (the distance between query embedding and the average embedding). Both cosine distance and mahalanobis distance are utilized in these algorithms.

The best-performing anomaly score is chosen for each machine type. It is noted that different anomaly scores are not aggregated into a comprehensive score since the performance of different scores varies greatly on different machine types. Once the anomaly score exceeds a predefined threshold, the clip is classified as anomalous. The threshold algorithm is the same with the official baselines \cite{Dohi_arXiv2022}.

One additional bonus of generative model is that it mirrors the input onto its learned knowledge and generates the reflected result, no matter it is based on deep understanding or trivial denoising. By comparing the original input and its reflection (reconstruction result), the difference can be localized. In the proposed AEGAN-AD model, the reconstructed anomalous spectrogram is compared with the mean spectrogram of the training set.

\section{Experiment}
\label{sec:experiment}

\subsection{Dataset}
The experiment is carried out on the dataset of DCASE 2022 Challenge TASK 2, a machine audio dataset where only normal clips are provided as training data. Domain shift is introduced to feature the variational working scenarios, where the majority of training data are from source domain. Detection performance is measured by AUC and pAUC. It is worthy to note that the calculation of AUC on either source domain or target domain takes all anomalous clips into account as required by DCASE competition rule. The performance of each baseline system is presented by the harmonic mean of all AUC and pAUC for each machine type. A harmonic mean over all machine types is also calculated to demonstrate the general performance.

\subsection{Model Settings}
In the AEGAN-AD framework, mel-spectrogram (2048-point FFT, 128 mel filters), taken logarithm, is calculated at first, then scales to [-1,1] by an affine transform, and sliced by a sliding window into overlapping segments of $128\times128$. Adam optimizers \cite{kingma2014adam} are adopted with a learning rate of 0.0002 for both networks. As for WGAN-GP parameters, $\lambda$ is chosen as 10 and $n_{critic}$ is 1. The model is trained for 60 epochs with a batchsize of 512. A set of parameters is saved for each machine type.

Seven baseline systems are selected as benchmarks in this work, including two official baselines, AnoGAN \cite{schlegl2017unsupervised}, GANomaly \cite{akcay2018ganomaly} and all three generative models from the top 15 submissions of the 2022 challenge, i.e. the work of Du \cite{DuNERCSLIP2022}, To\v{z}i\v{c}ka \cite{TozickaNSW2022} and Yamashita \cite{YamashitaGU2022}. AnoGAN and GANomaly are implemented mostly following the default settings.

\begin{figure*}[ht]
    \subfloat[All samples of training set]{
        \includegraphics[width=0.32\linewidth]{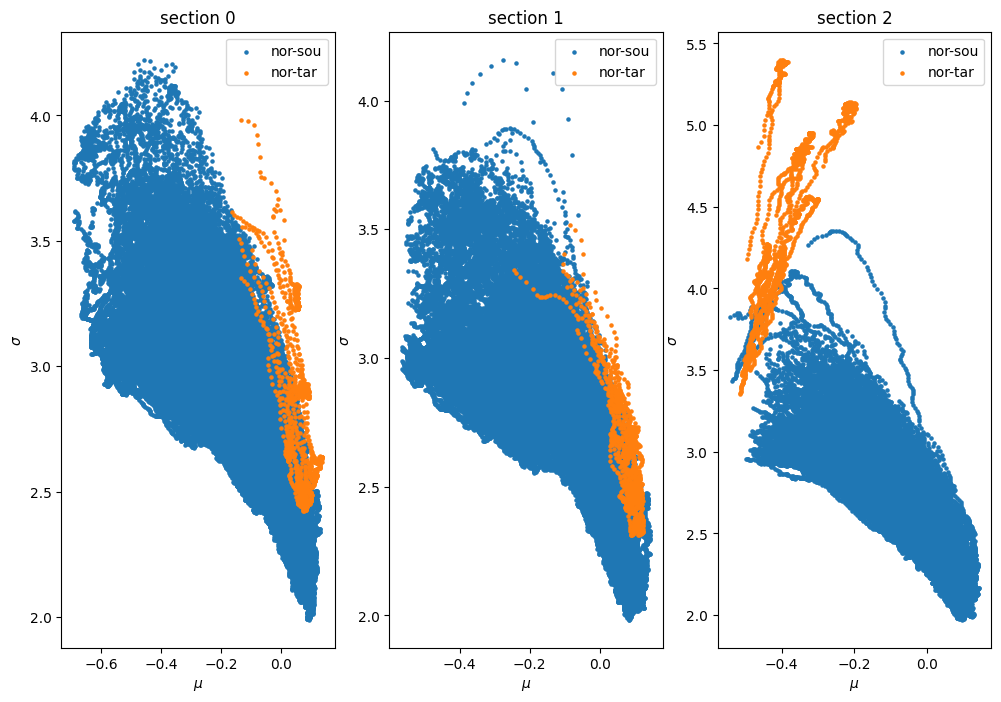}
        \label{fig:abla_dev}}
    \hfill
    \subfloat[Normal samples of test set]{
        \includegraphics[width=0.32\linewidth]{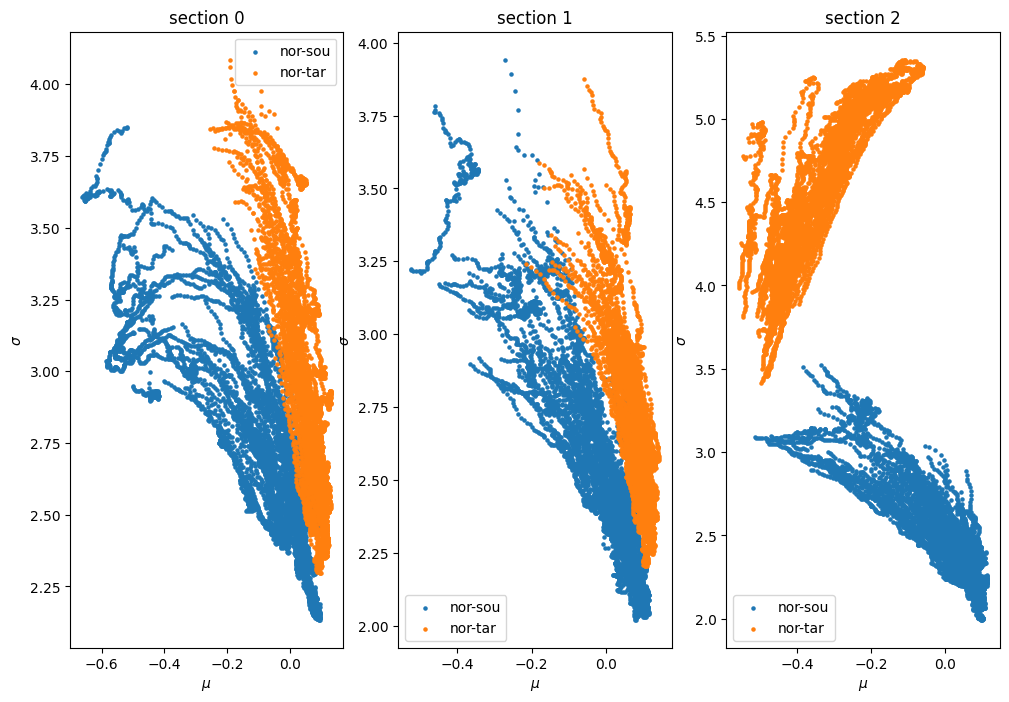}
        \label{fig:abla_test_nor}}
    \hfill
    \subfloat[All samples of test set]{
        \includegraphics[width=0.32\linewidth]{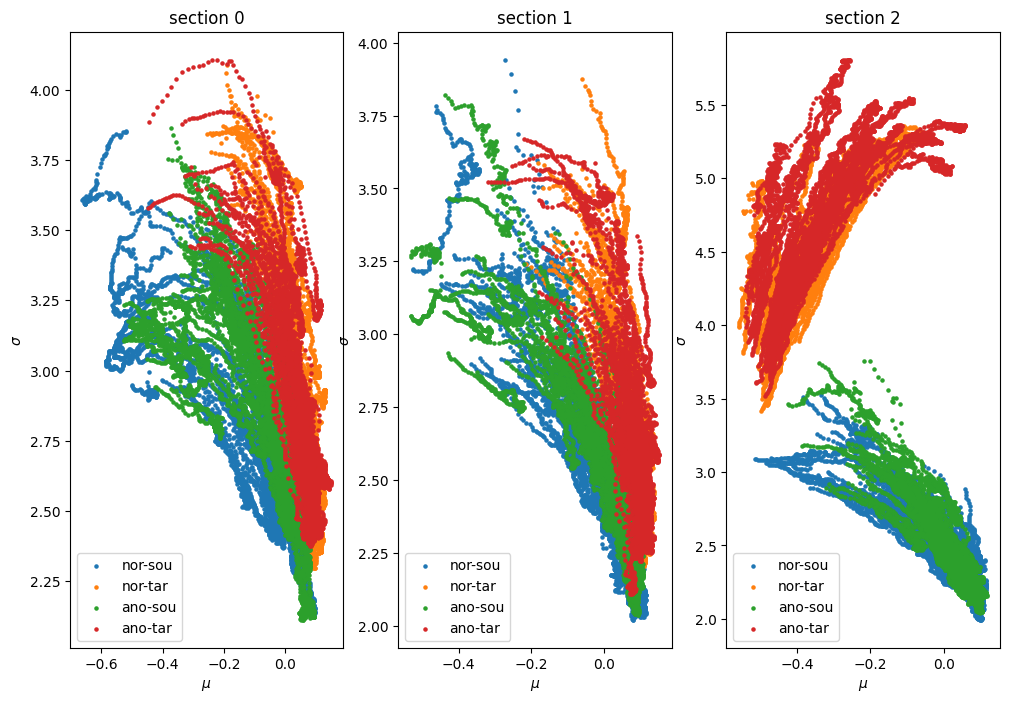}
        \label{fig:abla_test_all}}
    \caption{T-sne visualization of $\mu(x)$ and $\sigma(x)$ calculated by LN layer on ToyCar. (a) Visualization of all samples on the training set. Blue nodes and yellow nodes denote source and target normal samples respectively. (b) Visualization of all normal samples on the test set. (c) Visualization of all samples on the test set. Green nodes and red nodes denote source and target anomalies respectively. It illustrates the degree of domain shift.}
    \label{fig:ln_anal}
\end{figure*}

\begin{figure}[ht]
    \subfloat[bearing]{
        \includegraphics[width=0.479\linewidth]{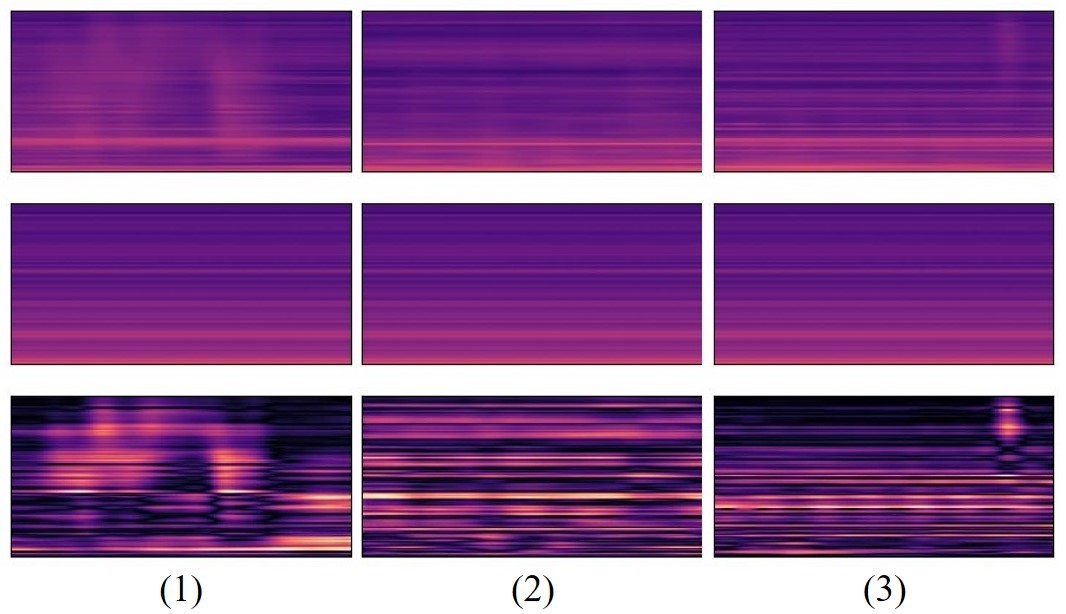}
        \label{fig:local_bearing}}
    \hfill
    \subfloat[ToyCar]{
        \includegraphics[width=0.479\linewidth]{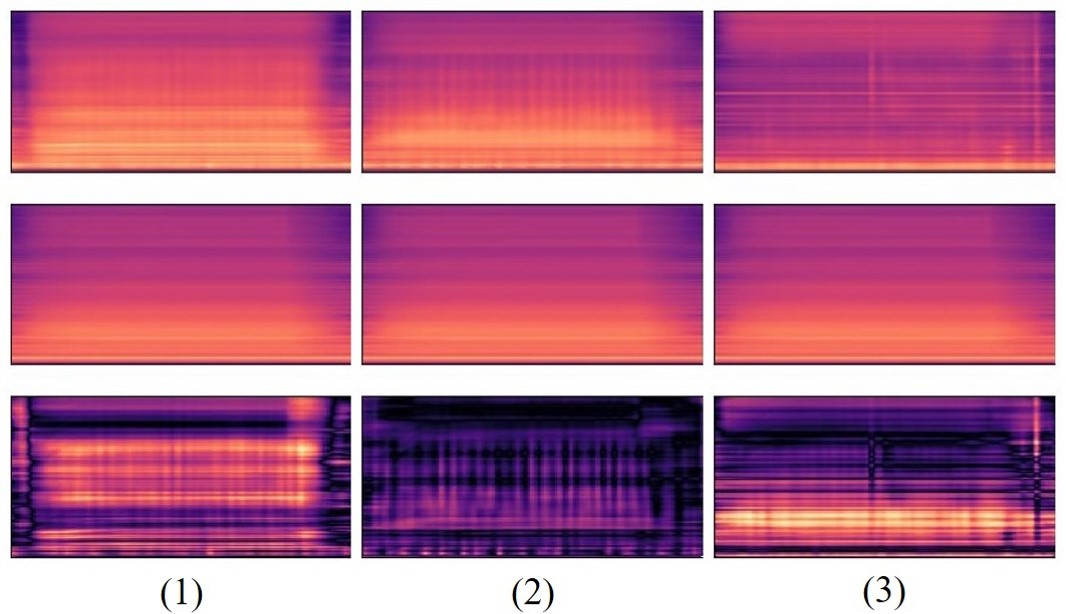}
        \label{fig:local_ToyCar}}
    \caption{Localization result of anomalous clips of bearing (a) and ToyCar (b). First row: Reconstructed query spectrogram. Second row: Mean spectrogram of the training set. Third row: Pixel-level localization resulting from calculating the difference between the above two rows. The brighter a pixel is, the more anomalous it appears to be.}
    \label{fig:local}
\end{figure}

\subsection{Performance Comparison}
Table \ref{tab:comp_related} presents the performance of AEGAN-AD along with seven baselines. AEGAN-AD surpasses all seven baselines on bearing, fan, gearbox, slider with an improvement of 10.91\%, 3.46\%, 1.19\% and 1.91\% respectively, while the autoencoder proposed by To\v{z}i\v{c}ka performs slightly better on ToyCar. AEGAN-AD is with the best performance among all generative models with a general improvement of 3.84\%, which demonstrates the superiority of our model. 

\subsection{Localization}
The localization results on bearing and ToyCar are presented in Fig \ref{fig:local}. For bearing, AEGAN-AD highlights the impulsive components and some improper frequency components; For ToyCar, AEGAN-AD highlights the absence of low frequency component and the periodic noise feature. In general, AEGAN-AD does capture how the query spectrogram deviates from the mean representation.

\subsection{Ablation Study}
It is believed that LN better handles domain shift than commonly used batch normalization (BN) \cite{ioffe2015batch}. To manifest this, a full LN model is compared with a hybrid model (BN for \textit{G} and LN for \textit{D} so as to meet the requirement of WGAN-GP). The performance is compared in Table \ref{tab:ablation}. In general, LN performs better than BN on most machine types except gearbox, which is believed to have minor difference between domains. As we dive deeper into the detailed performance on different domains, we observe that BN performs better on source domain, yet much worse on target domain. This suggests that normalizing within samples is better when training data is insufficient, yet when it is sufficient, normalizing within batch yields better result.

\begin{table}
    \centering
    \caption{Performance Comparison of Different Normalization Functions}
    \begin{tabular}{cccccc}
        \toprule
         Model & bearing & fan & gearbox & slider & ToyCar \\
         \midrule
         BN+LN & 63.23 & 64.62 & \textbf{75.27} & 71.16 & 70.79 \\
         LN & \textbf{76.03} & \textbf{65.83} & 71.94 & \textbf{74.06} & \textbf{78.46} \\
         \bottomrule
    \end{tabular}
    \label{tab:ablation}
\end{table}

To demonstrate how domain shift interferes training, the statistics of each LN layer are extracted and visualized by t-sne in Fig \ref{fig:ln_anal}. Comparison of Fig \ref{fig:abla_dev} with Fig \ref{fig:abla_test_nor} indicates that the minority of target domain makes it hard to infer the overall features of target domain when training. Comparison of Fig \ref{fig:abla_test_nor} with Fig \ref{fig:abla_test_all} reveals that the difference between domains is bigger than the difference between normal and anomalous samples.

\section{Conclusion}
\label{sec:conclusion}
In this article, we propose an unsupervised approach for machine audio anomaly detection utilizing GAN. We figure out that the advantages of GAN suit the need of audio anomaly detection, yet the denoising nature of reconstruction deprecates the performance. Therefore, the discriminator is redesigned to aid the generator in both training stage, i.e. promoting deep understanding, and detection stage, i.e. complementary perspectives. The experimental result on DCASE dataset surpasses all seven models generally, demonstrating the superiority of our AEGAN-AD model. A novel localization mechanism is also investigated. The ablation study indicates that LN is better at handling insufficient data.

\vfill\pagebreak

\bibliographystyle{IEEEbib}
\bibliography{refer.bib}

\begin{thebibliography}{10}

\bibitem{ramaswamy2000efficient}
Sridhar Ramaswamy, Rajeev Rastogi, and Kyuseok Shim,
\newblock ``Efficient algorithms for mining outliers from large data sets,''
\newblock in {\em Proceedings of the 2000 ACM SIGMOD international conference
  on Management of data}, 2000, pp. 427--438.

\bibitem{breunig2000lof}
Markus~M Breunig, Hans-Peter Kriegel, Raymond~T Ng, and J{\"o}rg Sander,
\newblock ``Lof: identifying density-based local outliers,''
\newblock in {\em Proceedings of the 2000 ACM SIGMOD international conference
  on Management of data}, 2000, pp. 93--104.

\bibitem{LiuCQUPT2022}
Ying Zeng, Hongqing Liu, Lihua Xu, Yi~Zhou, and Lu~Gan,
\newblock ``Robust anomaly sound detection framework for machine condition
  monitoring,''
\newblock Tech. {R}ep., DCASE2022 Challenge, July 2022.

\bibitem{goodfellow2020generative}
Ian Goodfellow, Jean Pouget-Abadie, Mehdi Mirza, Bing Xu, David Warde-Farley,
  Sherjil Ozair, Aaron Courville, and Yoshua Bengio,
\newblock ``Generative adversarial networks,''
\newblock {\em Communications of the ACM}, vol. 63, no. 11, pp. 139--144, 2020.

\bibitem{schlegl2017unsupervised}
Thomas Schlegl, Philipp Seeb{\"o}ck, Sebastian~M Waldstein, Ursula
  Schmidt-Erfurth, and Georg Langs,
\newblock ``Unsupervised anomaly detection with generative adversarial networks
  to guide marker discovery,''
\newblock in {\em International conference on information processing in medical
  imaging}. Springer, 2017, pp. 146--157.

\bibitem{akcay2018ganomaly}
Samet Akcay, Amir Atapour-Abarghouei, and Toby~P Breckon,
\newblock ``Ganomaly: Semi-supervised anomaly detection via adversarial
  training,''
\newblock in {\em Asian conference on computer vision}. Springer, 2018, pp.
  622--637.

\bibitem{DuNERCSLIP2022}
Du~Jun, Liu Diyuan, Wang Yajian, Wang Shuxian, Chu Fan, Li~Yunqing, Pan Jia,
  Wang Qing, and Gao Tian,
\newblock ``Ensemble of multiple anomaly detectors under domain generalization
  conditions,''
\newblock Tech. {R}ep., DCASE2022 Challenge, July 2022.

\bibitem{YamashitaGU2022}
Junya Yamashita, Ryosuke Tanaka, Keisuke Ikeda, Shiiya Aoyama, Satoshi Tamura,
  and Satoru Hayamizu,
\newblock ``Anomaly detection using autoencoder, idnn and u-net using
  ensemble,''
\newblock Tech. {R}ep., DCASE2022 Challenge, July 2022.

\bibitem{gulrajani2017improved}
Ishaan Gulrajani, Faruk Ahmed, Martin Arjovsky, Vincent Dumoulin, and Aaron~C
  Courville,
\newblock ``Improved training of wasserstein gans,''
\newblock {\em Advances in neural information processing systems}, vol. 30,
  2017.

\bibitem{salimans2016improved}
Tim Salimans, Ian Goodfellow, Wojciech Zaremba, Vicki Cheung, Alec Radford, and
  Xi~Chen,
\newblock ``Improved techniques for training gans,''
\newblock {\em Advances in neural information processing systems}, vol. 29,
  2016.

\bibitem{MIMII}
Kota Dohi, Tomoya Nishida, Harsh Purohit, Ryo Tanabe, Takashi Endo, Masaaki
  Yamamoto, Yuki Nikaido, and Yohei Kawaguchi,
\newblock ``{MIMII DG}: Sound dataset for malfunctioning industrial machine
  investigation and inspection for domain generalization task,''
\newblock {\em In arXiv e-prints: 2205.13879}, 2022.

\bibitem{ToyADMOS}
Noboru Harada, Daisuke Niizumi, Daiki Takeuchi, Yasunori Ohishi, Masahiro
  Yasuda, and Shoichiro Saito,
\newblock ``{ToyADMOS2}: Another dataset of miniature-machine operating sounds
  for anomalous sound detection under domain shift conditions,''
\newblock in {\em Proceedings of the 6th Detection and Classification of
  Acoustic Scenes and Events 2021 Workshop (DCASE2021)}, Barcelona, Spain,
  November 2021, pp. 1--5.

\bibitem{deng2022ens}
Yufeng Deng, Anbai Jiang, Yuchen Duan, Jitao Ma, Xuchu Chen, Jia Liu, Pingyi
  Fan, Cheng Lu, and Wei-Qiang Zhang,
\newblock ``Ensemble of multiple anomalous sound detectors,''
\newblock in {\em Proceedings of the 7th Detection and Classification of
  Acoustic Scenes and Events 2022 Workshop (DCASE2022)}, Nancy, France,
  November 2022.

\bibitem{radford2015unsupervised}
Alec Radford, Luke Metz, and Soumith Chintala,
\newblock ``Unsupervised representation learning with deep convolutional
  generative adversarial networks,''
\newblock {\em arXiv preprint arXiv:1511.06434}, 2015.

\bibitem{ba2016layer}
Jimmy~Lei Ba, Jamie~Ryan Kiros, and Geoffrey~E Hinton,
\newblock ``Layer normalization,''
\newblock {\em arXiv preprint arXiv:1607.06450}, 2016.

\bibitem{Dohi_arXiv2022}
Kota Dohi, Keisuke Imoto, Noboru Harada, Daisuke Niizumi, Yuma Koizumi, Tomoya
  Nishida, Harsh Purohit, Takashi Endo, Masaaki Yamamoto, and Yohei Kawaguchi,
\newblock ``Description and discussion on {DCASE} 2022 challenge task 2:
  Unsupervised anomalous sound detection for machine condition monitoring
  applying domain generalization techniques,''
\newblock {\em In arXiv e-prints: 2206.05876}, 2022.

\bibitem{kingma2014adam}
Diederik~P Kingma and Jimmy Ba,
\newblock ``Adam: A method for stochastic optimization,''
\newblock {\em arXiv preprint arXiv:1412.6980}, 2014.

\bibitem{TozickaNSW2022}
Jan Tozicka, Bezusek Marek, Durkota Karel, and Linda Michal,
\newblock ``Dadaed - double anomaly detector with aediff,''
\newblock Tech. {R}ep., DCASE2022 Challenge, July 2022.

\bibitem{ioffe2015batch}
Sergey Ioffe and Christian Szegedy,
\newblock ``Batch normalization: Accelerating deep network training by reducing
  internal covariate shift,''
\newblock in {\em International conference on machine learning}. PMLR, 2015,
  pp. 448--456.

\end{thebibliography}

\end{document}